# Spectral Estimation Methods: Comparison and Performance Analysis on a Steganalysis Application

Tolga Mataracıoğlu[1] and Ünal Tatar[2]

[1]Department of Electrical and Electronics Engineering, Middle East Technical University,
06531 Ankara, Turkey,
[2]Department of Mathematics and Institute of Applied Mathematics, Middle East Technical University,
06531 Ankara, Turkey,
{e157307,e156699}@metu.edu.tr

*Abstract* - **Steganography is the art and science of writing hidden messages in such a way that no one apart from the intended recipient knows of the existence of the message. In today's world, it is widely used in order to secure the information. In this paper, the traditional spectral estimation methods are introduced. The performance analysis of each method is examined by comparing all of the spectral estimation methods. Finally, from utilizing those performance analyses, a brief pros and cons of the spectral estimation methods are given. Also we give a steganography demo by hiding information into a sound signal and manage to pull out the information (i.e, the true frequency of the information signal) from the sound by means of the spectral estimation methods.**

*Index Terms* - **Steganography, Digital Signal Processing, Spectral estimation methods, The Periodogram Method, Blackman and Tuckey Method, Capon's Method, Yule-Walker Method, The Modified Covariance Method, Bartlett Window, Parzen Window.**

## I. INTRODUCTION

Steganography comes from the Greek word meaning "covered writing." The key concept behind steganography is that the message to be transmitted is not detectable to the casual eye. In fact, people who the are not intended to be the recipients of the message should not even suspect that a hidden message exists.

The difference between steganography and cryptography is that in cryptography, one can tell that a message has been encrypted, but he cannot decode the message without knowing the proper key [7][8]. In steganography, the message itself may not be difficult to decode, but most people would not detect the presence of the message. When combined, steganography and cryptography can provide two levels of security. Computer programs exist which encrypt a message using cryptography, and hide the encryption within an image using steganography.

Recently, computerized steganography has become popular. Using different methods of encoding, secret messages can be hidden in digital data, such as .bmp or .jpg images, .wav audio files, or e-mail messages.

In order to estimate the power spectra of the signals in Additive White Gaussian Noise, there exists some estimation methods [1]. Some of those are The Periodogram Method, The Blackman and Tuckey Method, Capon's Method, Yule-Walker Method, and Modified Covariance Method [2][4]. In this paper, all these spectrum estimation methods are examined and the performances of each are compared. Steganography is the art and science of writing hidden messages in such a way that no one apart from the intended recipient knows of the existence of the message [9]. In part II, the theoretical guidelines for those methods are given. In part III, simulation results and performance analyses of the methods are given. Also a steganography demo is given by hiding information into a sound signal and succeeded in pulling out the information by determining the true frequency of the information signal from the sound. Finally, a comment on simulation results, advantages and disadvantages of the spectrum estimation methods are given in part IV, the conclusion section.

## II. THEORETICAL GUIDELINES FOR SPECTRAL ESTIMATION METHODS

In this part, the power spectrum equations of each examined method are given briefly.

For the Periodogram Method, the equations of spectrum estimation and autocorrelation function are given below:

$$\hat{P}_{Per}(f) = \sum_{k=-(N-1)}^{N-1} \hat{r}_{xx}(k) e^{-j2\pi f k}$$

where

$$\hat{r}_{xx}(k) = \frac{1}{N} \sum_{n=0}^{N-1-|k|} x(n) x(n+|k|)$$

Here, N is the number of samples. One of the first uses of the periodogram spectral method, has been determining possible hidden periodicities in time series.

For the Blackman and Tuckey Method, the equations of spectrum estimation, autocorrelation function, and the window equation are given below:

$$\hat{P}_{BT}(f) = \sum_{k=-M}^{M} w(k)\hat{r}_{xx}(k)e^{-j2\pi fk}$$

$$\hat{r}_{xx}(k) = \begin{cases} \frac{1}{N}\sum_{n=0}^{N-1-k} x^*(n)x(n+k), k = 0,1,...,N-1 \\ \hat{r}_{xx}(-k), k = -(N-1),...,-1 \end{cases}$$

$$w_{Bartlett}(k) = \begin{cases} 1 - \frac{|k|}{M}, |k| \leq M \\ 0, |k| > M \end{cases}$$

In this method, smoothing by the spectral window will also have the undesirable effect of reducing the resolution.

For Capon's Method, the equation of spectrum estimation is given below:

$$S_{Capon}(f) = \frac{1}{e^T R_{xx}^{-1} e}$$

where

$$e = \begin{bmatrix} 1 \\ e^{-jw_0} \\ . \\ . \\ . \\ e^{-j(M-1)w_0} \end{bmatrix}$$

For Yule-Walker Method [6], the equation of spectrum estimation is given below:

$$S_{YW}(f) = \frac{\hat{\rho}}{\left|1 + \sum_{k=1}^{p} \hat{a}(k)e^{-j2\pi fkT}\right|^2}, k = 1,2,...,n$$

$$\hat{\rho} = \hat{R}_{xx}(0) + \sum_{k=1}^{p} \hat{a}(k)\hat{R}_{xx}(-k)$$

Finally, for Modified Covariance Method, the equation of spectrum estimation is given below:

$$S_{CM}(f) = \frac{\hat{\rho}}{\left|1 + \sum_{k=1}^{p} \hat{a}(k)e^{-j2\pi fkT}\right|^2}$$

$$\hat{\rho} = \hat{C}_{xx}(0,0) + \sum_{k=1}^{p} \hat{a}(k)C_{xx}(0,k)$$

To compare, it can be said that the Blackman-Tuckey estimates perform better than the corresponding Periodogram estimates. This means that the variance in Blackman-Tuckey estimate curves is smaller than that of the Periodogram estimate curves. In Blackman-Tuckey method, the noise level increases with M. Also the bias decreases when M increases.

### III. SIMULATION RESULTS

In this part, some popular spectral estimation methods are examined by using MATLAB.

a) Let our signal be,

$$x(n) = A\cos(2\pi f_1 n) + B\cos(2\pi f_2 n) + w(n)$$

where n=0,…,N-1, N=128, w(n) is AWGN with $10^{-3}$ variance.

i) $A = 1, B = 1, f_1 = 0.2, f_2 = 0.25, M = 5$

For this case, the estimation of the frequencies is given in Figure 1.

It's easily seen that the Modified Covariance Method gives the two frequencies exactly without any shift. After that, Conventional Capon's method comes. None of the methods except the Modified Covariance Method, gives the position of the frequencies. They can not separate the two adjacent frequencies.

ii) $A = 1, B = 1, f_1 = 0.2, f_2 = 0.22, M = 10$

Let's increase the order and make it twice. Also make the frequencies more adjacent to each other. For this case, the figure of the estimators is given in Figure 2.

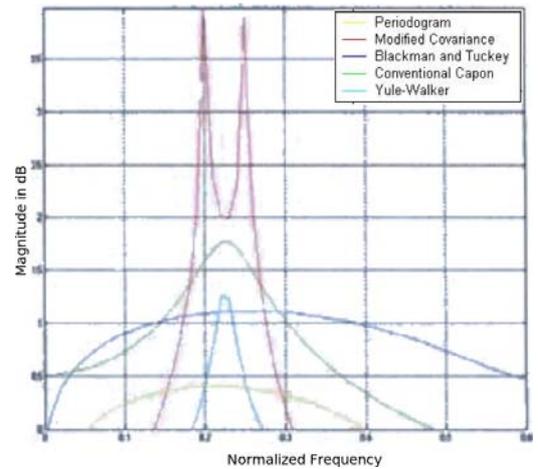

**Fig. 1.** Estimation of the frequencies for case a-i

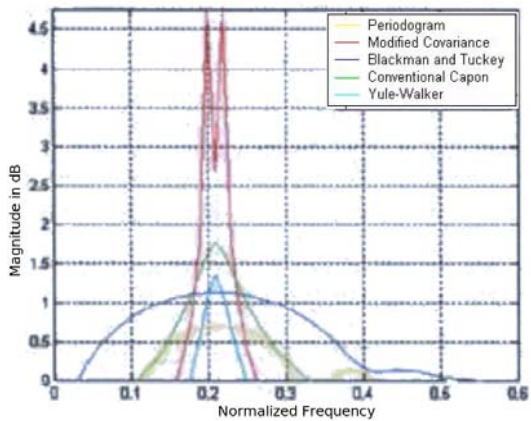

**Fig. 2.** Estimation of the frequencies for case a-ii

As the order increases, the variances of each estimator become smaller. The Modified Covariance Method is still the best. It finds the two adjacent frequencies exactly. However, other methods can not estimate both frequencies since their resolutions are much more less than the Modified Covariance Method's.

iii) $A = 1, B = 0.1, f_1 = 0.2, f_2 = 0.25, M = 10$

Now make the magnitude of the second sinusoid smaller. For this case, the figure obtained is given in Figure 3.

The Modified Covariance estimator still finds the two adjacent frequencies. Of course, for the second sinusoid, the magnitude of the estimator is smaller than the first one.

b) Suppose that we have only the autocorrelation function R, instead of the original signal. So we do not have to estimate the autocorrelation functions for some estimators. In addition, for those methods, the error has to be smaller and the estimators give better performance. For this case, only Blackman and Tuckey Method, Capon's Method, and the Yule-Walker Method are examined. Let our autocorrelation function be:

$$R(n) = A\cos(2\pi f_1 n) + B\cos(2\pi f_2 n) + \delta(n)$$

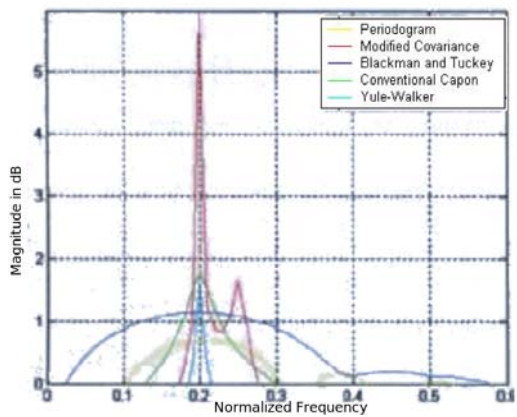

**Fig. 3.** Estimation of the frequencies for case a-iii

i) $A = 5, B = 5, f_1 = 0.2, f_2 = 0.3, M = 5$

For the given parameters, the related figure is given in Figure 4.

There is no doubt that the worst estimator is Blackman and Tuckey estimator as seen in the figure. Since the autocorrelation matrix is exact, the Yule-Walker Method and Capon's Method estimate the two frequencies with the highest resolution. However, if we compare the magnitudes, Yule-Walker estimator gives better performance than the Capon's estimator.

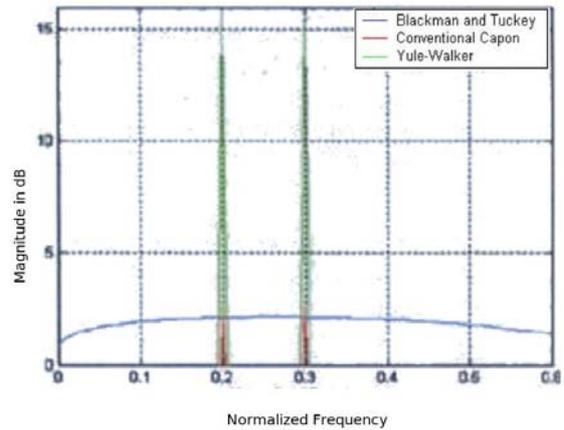

**Fig. 4.** Estimation of the frequencies for case b-i

ii) $A = 5, B = 5, f_1 = 0.2, f_2 = 0.3, M = 10$

So as to monitor the change in the performance, let us increase the order and make it twice. The other parameters remain constant. The figure obtained for this case is given in Figure 5.

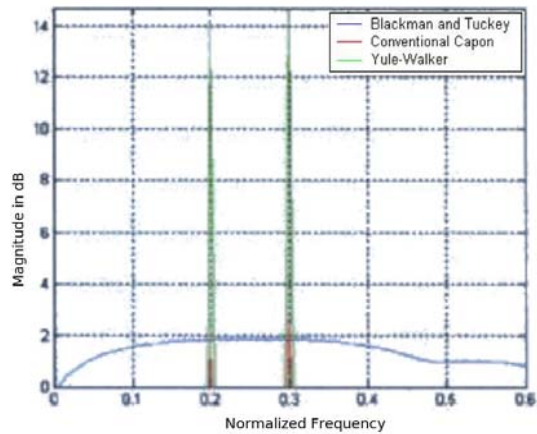

**Fig. 5.** Estimation of the frequencies for case b-ii

Since the two estimators found the two frequencies in the prior case, we can not see any improvement in this simulation. Also for the Blackman and Tuckey estimator, an extra ripple is seen since we increase the order.

c) In this case, a real data is concerned to analyse the practical situation. This data is a sound data and it is transmitted to the MATLAB by means of the Standard MATLAB function given below:

signal=wavread('sample.wav');

After performing this line, the original sound is transmitted to a vector. To give sound to the data vector, the standard MATLAB function given below is necessary:

sound(signal);

The waveform of the sound used for this paper is given in Figure 6.

Since transmitting the data into a vector, the length of the vector becomes too large. In order to make the simulations more rapid, the length of the data is ranged with a thousand. For this case, a sinusoid is added to the sound. This means that the sound is now the noise component of the signal. It's not white Gaussian [5]. Also the elements are correlated to each other.

  i) p=10

Let our signal be:

$$x(n) = \cos(0.4\pi n) + sound(n)$$

The figure obtained for this case is given in Figure 7.

It's easily seen from the figure that all the methods give worse performance since the noise part of the signal (here, the sound) is correlated and its shape is not Gaussian distributed. However the Modified Covariance Method estimates the frequency although its variance becomes larger.

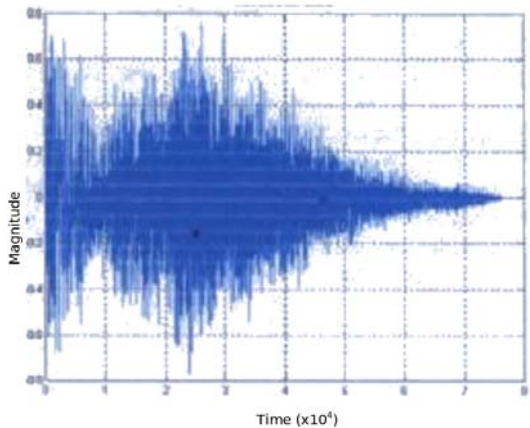

**Fig. 6.** The waveform of the sound used for this paper

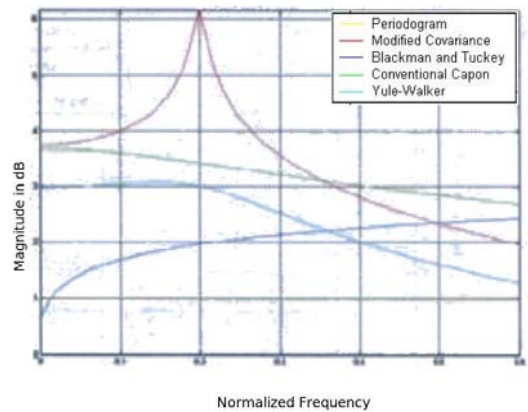

**Fig. 7.** Estimation of the frequencies for case c-i

ii) p=20

For more accurate estimations, let us increase the order and make it twice. The related figure is given in Figure 8.

It can be observed that the Yule-Walker Method begins to estimate the frequency, but the order is still not enough. If we remember the chosen order for white Gaussian noise cases, making the order twice is not enough since the sound data is correlated.

By means of these two simulations, we can say that we can add our information signal to a real data and send it to the receiver through a channel and the receiver takes the data and applies the best estimator to pull out the information from the real data. Steganography is the science of hiding information into another signal. This demo is an example of hiding information into a sound signal. So a sound steganography analysis has been performed in this paper.

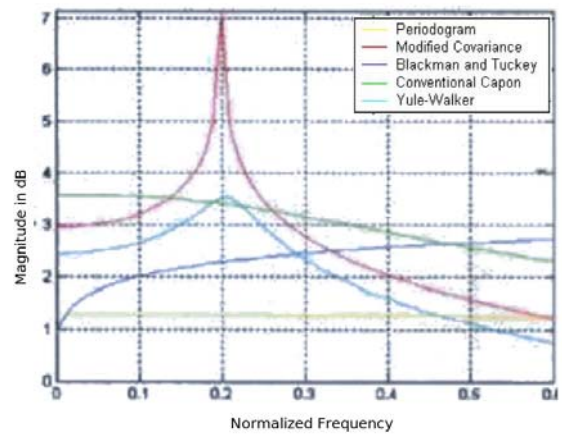

**Fig. 8.** Estimation of the frequencies for case c-ii

## IV. CONCLUSION

In this paper, we discussed, examined, and compared the performances of the traditional spectral estimation methods by using steganalysis of a sound signal. In the introduction part and in the second part, we gave the theory of the estimators.

After that, we simulated the methods and found out the best one for different cases.

To summarize the whole paper, we'll give the advantages and the disadvantages of the spectrum estimation methods. It's obvious that all spectrum estimators show different performances for different cases. The rest of the paper deals with this phenomenon and the results of the application of the sound steganography, i.e, hiding information into a sound signal.

If we have a deterministic signal (i.e, a pure sinusoid), then the non-parametric methods such as the Periodogram Method and the Blackman and Tuckey Method show better performances than the autocorrelation based methods such as Capon's Method, Yule-Walker Method, and the Modified Covariance Method.

However, if we want to form a stochastic signal, we briefly add white Gaussian noise to our original signal, i.e, a sinusoid. For this case, autocorrelation based methods such as Capon's Method, Yule-Walker Method, and the Modified Covariance Method show better performances than the non-parametric methods.

For the non-parametric methods, we can conclude that they show better performances as the length of the data increases [3]. However, we can not say the same words as the length of the data becomes smaller. Also their simulations last short since the codes of those are simple.

In contrast, the non-parametric methods do not show very good performances for stochastic signals. Also since we use windowing for the Blackman and Tuckey Method, the spectrum can give negative values if we do not use a proper window such as Bartlett Window or Parzen Window.

For the other examined methods, it can be easily said that the size of the filter (also called order) determines the performances of the estimators. If it is large enough, the method estimates all frequencies exactly. However, if the order of the method is too large, then extra peaks appear in the spectrum. The simulations are now not very simple since we have to find the parameters. So the code becomes more complex. If we use the exact autocorrelation function instead of using the estimate, then the autocorrelation based spectral estimation methods show the excellent performances while the other methods show poor performances.

So as to analyse the situation of not using additive white Gaussian noise, we used a sound data. Since its elements are correlated to each other and the waveform of the sound data is not Gaussian shaped, we can add this data to our sinusoid and behave this signal like a sinusoid with correlated noise. This means that we hide the information signal in the sound. This method is called steganography. The results are not astonishing. All the spectral estimation methods show worse performances if we compare the results with AWGN case. The best performance is shown by the Modified Covariance Method. However, the variance of this method's spectrum becomes larger.

As a result, the Modified Covariance Method shows the best performance for any case, performed in this paper, among the examined spectral estimation methods. This method needs quite small order to estimate the frequencies. Also the magnitude of the different sinusoids do not effect the performance of the estimator.

In addition, this method can separate the frequencies which are very close to each other. If it begins to give worse performance while making the frequencies closer, then just increasing the order a little bit will be enough to separate the frequencies. Of course, this method shows excellent performance when the autocorrelation matrix is exact. Nevertheless, this method's performance of estimating the autocorrelation matrix is quite good. Also, this method is the best for the signals in noise. In addition, the type of the noise (i.e, AWGN or correlated and not Gaussian shaped noise) does not effect the performance of this method too much. Only the variance of the estimator becomes larger. In contrast, this method shows poor performance when the noise is dismissed. This means that, for deterministic signals, the Modified Covariance Method must not be chosen for estimation.

In our future work, the analysis of sound steganography (i.e, hiding information signal into a sound signal) will be made when the transmission channel is Rician and the source data is coded with low density parity check (LDPC) coding. Also beside using binary shift keying (BPSK), other modulation types will be combined with LDPC codes for the coded-modulation type of the system.